\documentclass[onecolumn,pra]{revtex4-2}
\usepackage{graphicx,psfrag,color}
\def\bc{\begin{center}}
\def\ec{\end{center}}

\def\beq{\begin{equation}}
\def\eeq{\end{equation}}

\def\d{\downarrow}
\def\u{\uparrow}

\def\bc{{\bf c}}

\def\prl{Phys. Rev. Lett.}

\def\prb{Phys. Rev. B}
\def\rmp{Rev. Mod. Phys.}

\begin{document}

\title{Monitoring photon entanglement in coupled cavities}

\author{Moises Acero$^1$, Jeremiah Harrington$^1$, Oleg L. Berman$^{1,2}$ and Klaus Ziegler$^{2,3}$}
\affiliation{
$^{1}$The Graduate School and University Center,\\ The City University of New York,
New York, NY 10016, USA \\
$^{2}$Physics Department, New York City College of Technology,\\
The City University of New York, Brooklyn, NY 11201, USA \\
$^3$   Institut f\"ur Physik, Universit\"at Augsburg,
D-86135 Augsburg, Germany}
\date{\today}

\begin{abstract}
We study the dynamics of $N$ photons in a Fock state, initially
located inside one cavity, and coupled by an optical fiber
to a second cavity. The entanglement of the photons is monitored
by projective measurements, repeated with a fixed time step.
This approach is applied to the formation of a photonic N00N state.
We calculate the probability of the transition of $N$ photons
from the left to the right cavity and the probability of the
return of $N$ photons to the left cavity under repeated projective 
measurements. The entanglement is analyzed for the N00N state
by its fidelity and its phase sensitivity, while for the entanglement 
between the states in the two cavities the entanglement entropy
is calculated. In addition, we study the monitored evolution of photons
in a single cavity, which are coupled to a single qubit, using the
Jaynes-Cummings model. Photon entanglement is analyzed in terms
of the entanglement entropy. In all these cases we find that entanglement
is sensitive to the details of monitoring protocol, which can be used
to control photon entanglement for specific applications.  
\end{abstract}

\maketitle

\section{Introduction} 

Photons are ideal candidates to study entanglement of quantum states, since
they have no interparticle interaction. This provides a platform in which thermal effects
do not play an essential role and an arsenal of different optical tools are available for
controlling and measurement~\cite{haroche13}.  
Since the concept of quantum entanglement was introduced, quantum
entanglement has been receiving more and more
attention~\cite{Nielsen,Horodecki}. Specifically in recent years,
with the interdisciplinary development of quantum informatics, the
general definition, physical properties and entanglement measurement
of quantum entanglement have been comprehensively and deeply
studied. Due to its nonlocality, quantum entanglement has been
widely used in quantum information, especially in quantum computing
and quantum communication~\cite{Zou}.

\begin{figure}[t]
\begin{center}
\includegraphics[width=10cm,height=7cm]{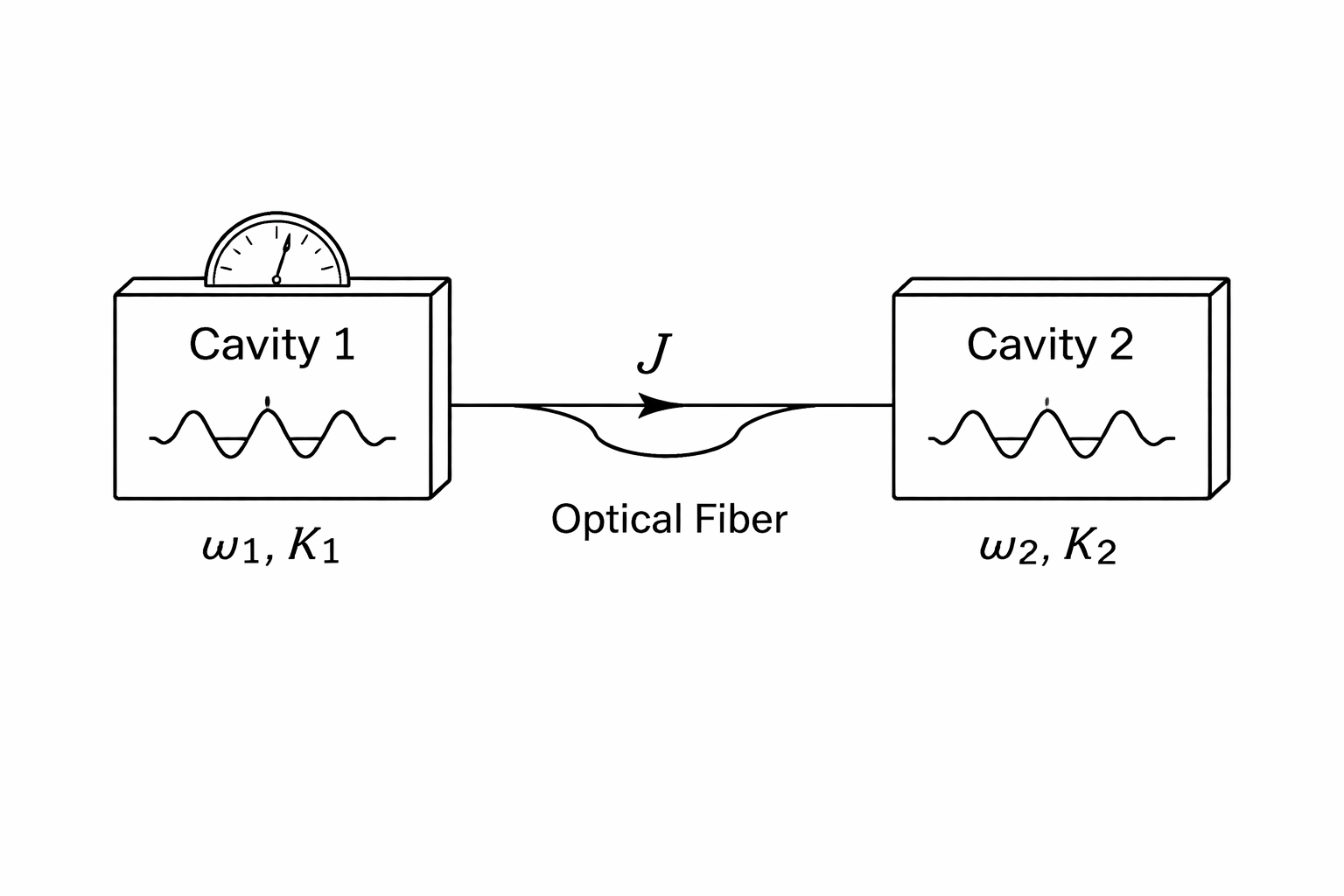}
\caption{Two optical cavities coupled by an optical fiber are subject to repeated
projective measurements in the left cavity.} 
\label{cavities}
\end{center}
\end{figure}

Although it is known that the effect of entanglement vanishes exponentially with
the number of photons, there might be ways to enhance entanglement by monitoring
with measurements. This was predicted some time ago in Ref.~\cite{kok02} by applying
projective measurements through beam splitters in a Mach-Zehnder interferometer.
Another option is to create an effective interaction of the photons by coupling them to
external degrees of freedom, such as two-level atoms.
A more recent proposal to create entanglement in multi-photon systems is to use the 
holonomy of an adiabatic change in a degenerate Hilbert space~\cite{szameit22}. 

Optical cavities play a similar role  for photons as the potential wells play for ultracold 
atoms~\cite{haroche13}. Photonic Fock states have been observed experimentally in an optical 
cavity~\cite{Brune08,Wang08}. The photons can interact indirectly with each
other via the interaction with atoms inside the optical cavities~\cite{Carmichael08,Schuster08,Ziegler_Laser}. 

When two optical cavities are coupled by an optical fiber superpositions of Fock states appear dynamically.
For instance, the initial Fock state $\left| N, 0 \right\rangle$ with all photons in the left cavity can
tunnel to the right cavity and vice versa. In this process the $N00N$ state $\left(\left| N, 0 \right\rangle 
+ \exp\left(i\phi N\right)\left|0, N \right\rangle \right)/\sqrt{2}$ can be formed during the evolution, which is
a highly entangled state. The $N00N$ state has been the subject of intensive studies, since it can
be applied for different precision measurements including very accurate
interferometry~\cite{Lee,Walther,Mitchell,Afek} and for optical lithography~\cite{Boto}. There were 
developed different methods to create  photonic $N00N$ states~\cite{kok02,kok02a,Cable}.   
In Ref.~\cite{kok02} it was demonstrated that conditioning the output of a linear optical
setup on projective measurements makes it possible to generate
path-entangled photon number states with more than two photons (up
to $4$ photons). The $N00N$ states up to $N = 5$ photons have been
created experimentally~\cite{Afek}.  
The creation of an eight-photon $N00N$ state with
genuine multipartite entanglement was observed in the
experiment~\cite{Yao12}. The experimental demonstration of quantum
entanglement among ten spatially separated single photons was
reported~\cite{Wang16}.   The experimental generation of the first
multi-photon entangled state where both the number of particles and
dimensions are greater than two has been demonstrated~\cite{Malik16}.
Two photons in this state reside in a three-dimensional space,
whereas the third lives in two dimensions.
Experimental quantum state tomography of $N00N$ states with up to four photons  was
performed~\cite{Afek2}. It was shown that a quantum polarized light
microscope, using entangled photon $N00N$ states with photon $N = 2$
and $N = 3$, can provide phase supersensitivity beyond the standard
quantum limit.  Therefore, the formation of the entangled $N00N$ states for the photons is 
possible~\cite{Israel}. A scheme in which $N00N$ states are the elementary resources for building 
quantum error correction codes against photon losses was also presented~\cite{Bergmann}.

There are several concepts to characterize entanglement: either by considering the entanglement
of two specific states in a high-dimensional Hilbert space or by probing the entanglement
entropy which characterizes the entanglement of two subspaces of the entire Hilbert
space. In the following we will use both concepts to study the entanglement of
$N$ photons in two coupled cavities. Moreover, we will investigate the entanglement
entropy of a qubit in a cavity with $N$ photons and how it is affected by measurements. 

The main motivation for this Paper is to provide an approach to
monitor the entanglement of photons. It is based on a measurement
protocol. We consider two photonic cavities, which are coupled by an
optical fiber that allows the photons to tunnel from one cavity to
the other. The results of this dynamics is entanglement of photonic
states. The dynamical entanglement of photons in the system of
coupled cavities was studied in Ref.~\cite{Klaus_JPCS}.
 In this Letter we study how the entangled photonic $N00N$ state can be 
 controlled by projective measurements, repeated
periodically with the time step $\tau$.

In this Paper we consider the Fock state of $N$ photons initially
located inside one cavity. Then we assume the first cavity to be
connected to the second cavity by a waveguide or a fiber. We
calculate the probability of transition of $N$ photons from the
first cavity to the second cavity $|c_{0}(t)|$ and the probability
of return of $N$ photons back to the first cavity $|c_{N}(t)|$ in
the presence of control by projective measurements, repeated
periodically with the time step $\tau$. To quantitatively measure
the entanglement between $\left| N, 0 \right\rangle$ and $\left| 0,
N \right\rangle$ Fock states one can use the probability
distribution $P\left(|c_{0}(t)|,|c_{N}(t)|\right)$ which measures
how often certain values of $|c_{0}(t)|$, $|c_{N}(t)|$  are visited
during the evolution in a period of time~\cite{Klaus_JPB}. Another
quantitative measure of the entanglement between $\left| N, 0
\right\rangle$ and $\left| 0, N \right\rangle$ Fock states can be
the Husimi $Q$ function~\cite{HarocheRaimond,ziegler17}. We have
demonstrated that while for the unitary evolution the entanglement
entropy is an oscillating function of time for $N = 2$ and $N = 20$
photons, if we consider measurements at the certain number of
measurements the EE can increase up to some
sufficient value for $N = 20$ photons.  In addition, we have shown
that another measure of the entanglement the difference $\Delta$ for
the $N00N$ states can be increased at the certain number of
measurements for $N = 10$ photons.

In addition to the coupled cavities, we consider  a photonic cavity
with one qubit inside. Then the cavity photons become entangled
through their interaction with this qubit. The resulting
entanglement is controlled by repeated projective measurements,
which are periodically repeated with the time step $\tau$.

We study the quantum entanglement created between cavity photons
coupled to one qubit applying a Jaynes-Cummings model. We employ the
Hamiltonian used for the Jaynes-Cummings
model~\cite{jaynes63,Knight} for one qubit, coupled to cavity
photons. While in the two coupled cavities the total number of
photons is conserved, the number of photons in the Jaynes-Cummings
model is not conserved since the qubit can absorb or emit a single
photon.

As a measure of quantum entanglement we use the R\'{e}nyi
EE, which  measures the quantum correlations between two subsystems under a spatial
bipartition~\cite{Bombelli,Srednicki,Eisert,Miao}, which has been
applied for detecting measurement-induced entanglement
transitions~\cite{Li,Skinner,Lunt}. Recently, the R\'{e}nyi
EE has been efficiently measured in the experiment~\cite{vanEnk,Elben,Satzinger}.

The paper is organized as follows: In Sect. \ref{sect:model} we define the model of two
photonic cavities, which are coupled by an optical fiber. Then the concept of photon 
entanglement in terms of the N00N state is discussed in Sect. \ref{sect:entanglement}.
Our measurement protocol for the monitored evolution is defined (Sect. \ref{sect:monitoring}). 
For the entanglement of the two cavities we introduce and calculate the R\'enyi entropy in 
Sect. \ref{sect:ent_ent}. The latter is also considered for the Jaynes-Cummings model 
under monitoring (\ref{sect:JC_model}). The results of the coupled cavities and the 
cavity with a single qubit are discussed and compared in terms of photon entanglement
(Sect. \ref{sect:discussion}) and some conclusions are presented in Sect. \ref{sect:conclusions}.

\begin{figure}[t]
\begin{center}
\includegraphics[width=1\linewidth]{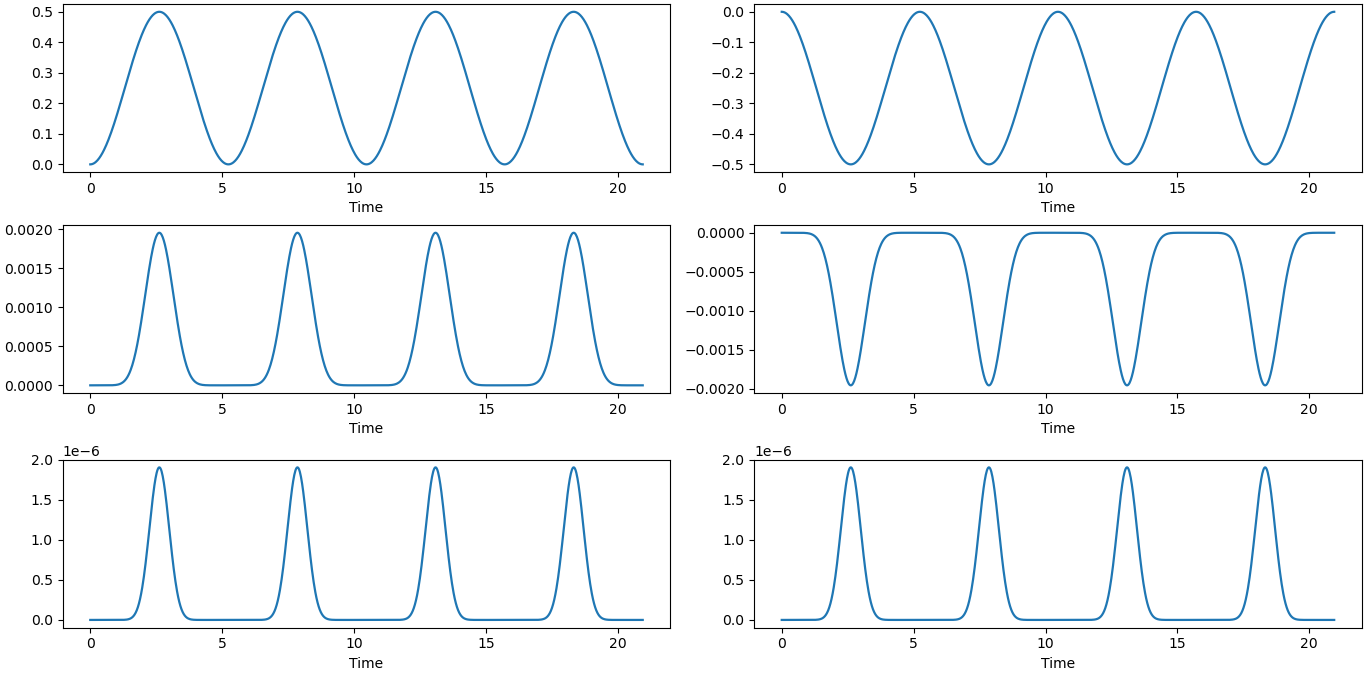}
\caption{
Unitary evolution: 
$p_e=2|c_0||c_N|$ (left column) and the difference $\Delta=2|c_0||c_N|\cos \phi$ 
of the probabilities of N00N states 
(right column) for $N=2,10,20$. The time is given in units of $\hbar/J$.
}
\label{fig:1}
\end{center}
\end{figure} 

\section{Model of two coupled cavities}
\label{sect:model}

The central idea is to prepare an eigenstate of the cavity as
initial state and then change the conditions of the system
by coupling another cavity through an optical fiber as shown in
Fig.~\ref{cavities}. As a result of this change, the system starts
to evolve in the Hilbert space to visit linear combinations of the
new Hamiltonian's eigenstates.
During a monitored evolution the system may create entangled states with
certain probability.
The dynamics and the entangled states will be calculated in
the following subsections for three different cases. In particular, we
will study the effect of the monitoring through repeated projective measurements
on the entanglement. 

The Hamiltonian of two uncoupled harmonic cavities is $
H_{hc}=\omega_0 \sum_{j=1,2} a^\dagger_j a_j $, where the index
$j=1,2$ refers to the cavity $j$, and the photon creation
(annihilation) operator $a_j^\dagger$ ($a_j$) acts in the cavity $j$.
This Hamiltonian has product Fock states
$|N-k,k\rangle\equiv|N-k\rangle|k\rangle$ ($k=0,..,N$) as
eigenstates. Next we consider the Hamiltonian of two coupled cavities:
\beq
\label{hamiltonian00} 
H_{hc}=-J(a_1^\dagger a_2+a_2^\dagger
a_1)+\omega_0 (a^\dagger_1 a_1+a^\dagger_2 a_2) 
.
\eeq 
Its eigenstates are
\beq
|E_k\rangle=\frac{2^{-N/2}}{\sqrt{k!(N-k)!}}(a_l^\dagger+a_r^\dagger)^k
(a_l^\dagger-a_r^\dagger)^{N-k}|0,0\rangle
\eeq
with equidistant eigenvalues $E_k=-J(N/2-k)$ ($k=0,1,...,N$).
Thus, the fastest oscillations in the unitary evolution occur with frequency $NJ/2$. 
Here we assume that $N$ is even and calculate the overlap of the eigenstates $|E_k\rangle$
with the Fock states of the uncoupled cavities.
First, the overlap of the initial Fock state $|N,0\rangle$ with the eigenstates of $H_{hc}$
is~\cite{ziegler12}: 
\beq
\label{binomial} 
\langle N,0|E_k\rangle
=2^{-N/2}{N\choose k}^{1/2} \ ,
\eeq 
which is non-zero for all eigenstates.
Thus, there is a binomial distribution for the spectral weight 
$|\langle N,0|N-k;k\rangle|^2$, which are transition probabilities for $k$ photons
moving to the other cavity. Its maximum is an equally distributed number of photons
over the two cavities. For large $N$ the binomial distribution becomes a Gaussian
distribution, where the width of the envelope is related to the energy level spacings 
$\Delta E=2J$. The Gaussian result resembles the Central Limit Theorem for independent
photons. Such a behavior was also found previously for freely
expanding bosons from an initial Fock state \cite{cramer08}. 

Eq. (\ref{binomial}) enables us to calculate the return amplitude $c_0(t)$ during the time $t$
and the transition amplitude $c_N(t)$ for all photons moved from the left to the right cavity
as
\beq
\label{amplitudes0}
c_0 =\langle N,0|e^{-iHt}|N,0\rangle
=\cos^N(Jt/2) , \ \ \
c_N =\langle 0,N|e^{-iHt}|N,0\rangle
=(-i)^N\sin^N(Jt/2)
\ .
\label{non_int_exp}
\eeq
For short times (i.e., for $J t\ll 1/\sqrt{N}$) we have a Gaussian decay of the Fock state
\beq
|c_0|=|\cos^N(Jt/2)|\sim e^{-J^2Nt^2/8}
\ .
\label{decay0}
\eeq
The velocity $d|c_0|/dt\sim -J^2Nt/4$ of the decay vanishes for $t\sim0$. Such a behavior
differs from the exponential decay of classical systems, and is related to the quantum 
Zeno effect \cite{misra77}.

\begin{figure}[h]
\begin{center}
\includegraphics[width=1\linewidth]{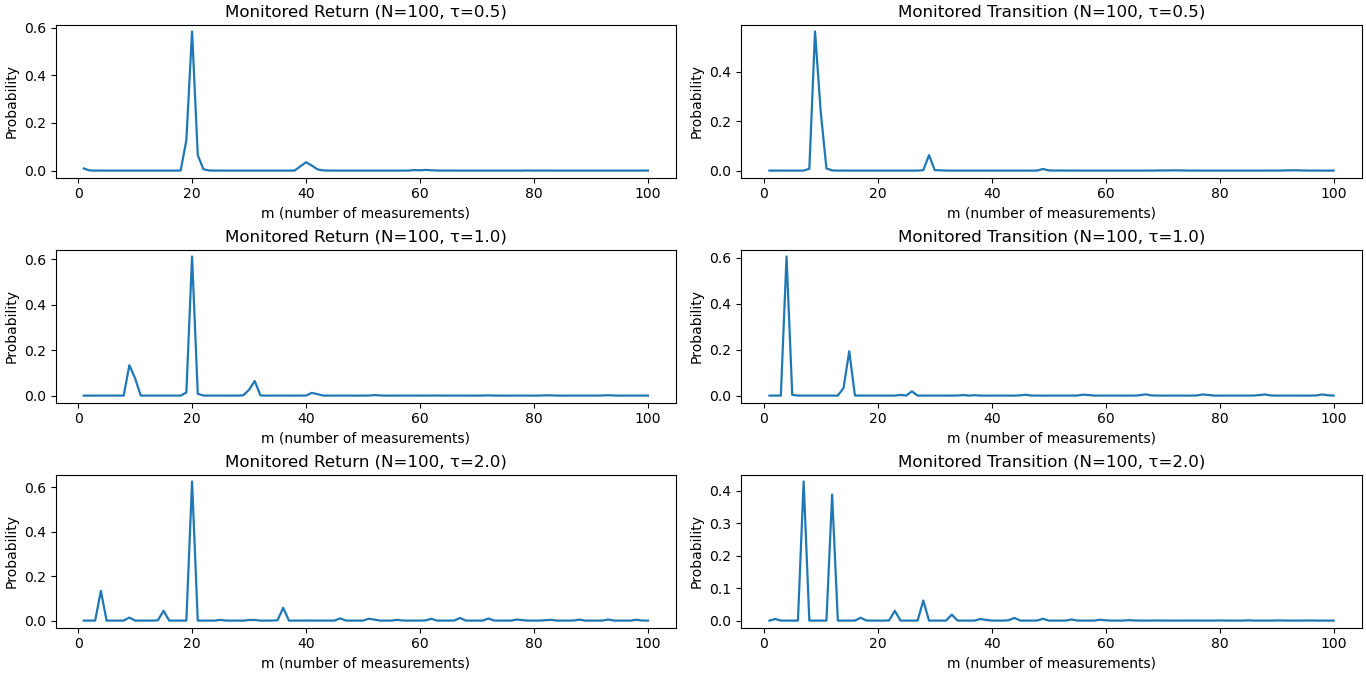}
\caption{
Monitored evolution:
Probabilities of the return to the initial state $|N,0\rangle$ (left) and the 
transition to the state $|0,N\rangle$ (right) for $N=100$.
}
\label{fig:2}
\end{center}
\end{figure} %

\subsection{Entanglement of photons}
\label{sect:entanglement}

In this Section we will focus on the evolution of the initial state, in which all photons are 
located in the left cavity as the product Fock state $|N,0\rangle$: 
$|\Psi_t\rangle=T_t|N,0\rangle$, where the unitary evolution operator $T_t$ is $e^{-iHt}$.
Then we determine the return amplitude to the initial state $c_0=\langle N,0|\Psi_t\rangle$ and the transition 
amplitude $c_N=\langle 0,N|\Psi_t\rangle$, which corresponds to the transition
of $N$ photons from the left to the right cavity. 
With these two amplitudes it is possible to calculate 
the overlap of $|\Psi_t\rangle$ with the N00N state
$|N00N\rangle=\left(|N,0\rangle +e^{i\phi N}|0,N\rangle\right)/\sqrt{2}$~\cite{sanders89} as
\beq
\langle N00N|\Psi_t\rangle
=\frac{c_0+e^{i\phi N} c_N}{\sqrt{2}}
\ .
\eeq
The phase $\phi$ of the N00N state can be determined by the expectation value of 
the operator $A=|N,0\rangle\langle 0,N|+|0,N\rangle\langle N,0|$ as~\cite{kok02}
\beq
\langle N00N|A|N00N\rangle=\cos \phi
\ .
\label{phase0}
\eeq
While the
evolution of the Fock state $|0,N\rangle$ will not create a pure N00N state, it creates
the state
\beq
|\Psi_t\rangle=c_0|N,0\rangle + c_N|0,N\rangle +\sum_{k=1}^{N-1}c_k|k,N-k\rangle \ , \ \ \ 
A|\Psi_t\rangle=c_N|0,N\rangle + c_0|N,0\rangle
\ ,
\eeq
which represents a pseudo N00N state (PNS) if $c_0,c_N\ne 0$ simultaneously. Then, 
instead of (\ref{phase0}), we get a phase-sensitive expression of the PNS as
\beq
\label{coherence0}
\langle\Psi_t|A|\Psi_t\rangle=2Re(c_0^* c_N)
=2|c_0c_N|\cos\Phi
.
\eeq
For a pure N00N state this expectation value of $A$ is also known as the fidelity
\beq
F=\frac{1}{2}|\langle N00N|\Psi_t\rangle|^2
=\frac{1}{2}\left[|c_0|^2+|c_N|^2+2|c_0||c_N|\cos(\Phi+\phi N)\right]
,
\label{husimi0}
\eeq
The fidelity is subject to the inequality $0\le|\langle N00N|\Psi_t\rangle|^2\le1$, 
which is $1/2$ at $t=0$ in the present case due to $|c_0(t=0)|^2=1$.
We note that the fidelity gives the Husimi--$Q$ function as $F/\pi$~\cite{HarocheRaimond}.

\begin{figure}[t]
\begin{center}
\includegraphics[width=1\linewidth]{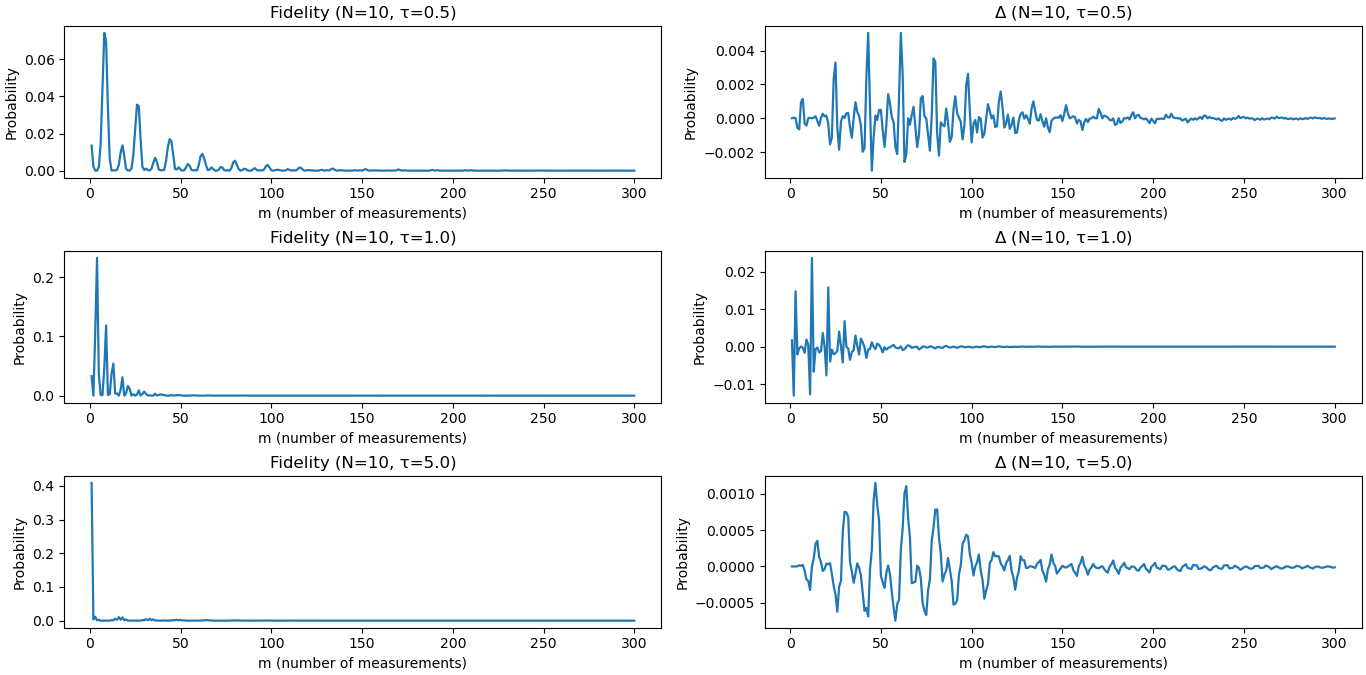}
\caption{
Probabilities of the monitored transition to a N00N state for $N=10$ photons: 
fidelity $|(\langle 0,N|+\langle N,0|)|\Psi_t\rangle|^2/2$ (left column) 
and the difference $\Delta$ of the N00N states (right column)
for different time steps $\tau=0.5, 1, 5$.
}
\label{fig:3}
\end{center}
\end{figure} 

Moreover, $A^2$ is the projector onto the space spanned by $\{|0,N\rangle, |N,0\rangle\}$: 
\[
A^2|\Psi_t\rangle=c_0|0,N\rangle + c_N|N,0\rangle
\ ,
\]
such that $\langle\Psi_t|A^2|\Psi_t\rangle=|c_0|^2+|c_N|^2$ and for $\phi=0$ we get
\beq
F=\frac{1}{2}\left(\langle\Psi_t|A^2|\Psi_t\rangle + \langle\Psi_t|A|\Psi_t\rangle \right)
\ .
\eeq
In the following we will study the product of the coefficients $|c_0c_N|$ as functions of 
time $t$ and the resulting expectation value $\langle\Psi_t|A|\Psi_t\rangle$. Moreover, 
to distinguish the PNS from the pure Fock states, 
we introduce the probability for the creation of a PNS and its maximum with respect to time
$P_e$ as 
\beq
p_e(t)=2|c_0c_N|
,\ \  
P_e=2\max_t|c_0c_N|
.
\label{ent00}
\eeq
Since the N00N state depends on the phase $\phi$, the fidelity depends on this phase
(cf. Eq. (\ref{husimi0})). The difference of the fidelities with $\phi=0$ and with
$\phi=\pi$ reads
\beq
\label{difference1}
\Delta=
\frac{1}{2}|\left(\langle N,0|\Psi_t\rangle + \langle0,N|\Psi_t\rangle\right)|^2
-\frac{1}{2}|\left(\langle N,0|\Psi_t\rangle - \langle0,N|\Psi_t\rangle\right)|^2
=2|c_0||c_N|\cos\Phi
.
\eeq
According to Eq. (\ref{coherence0}), $\Delta$ can also be written as  
$\langle\Psi_t|A|\Psi_t\rangle$. Subsequently, we will study $F$, $\Delta$ and $p_e$.

We start with the unitary evolution with $T_t=e^{-iHt}$
the measure for entanglement $p_e(t)=2|c_0c_N|$ reads with the amplitudes given in Eq. (\ref{amplitudes0})
\beq
p_e(t) =\frac{|\sin^N(Jt)|}{2^{N-1}}
\ ,
\label{ent1}
\eeq
such that the maximum with respect to time is
\beq
P_e=2^{-(N-1)}
.
\label{ent0}
\eeq
This indicates an exponential decay of the entanglement with the number of photons,
which is also reflected for the coherent part of the fidelity (\ref{husimi0})
\beq
F=\frac{1}{2}\left[\cos^{2N}(Jt/2)+\sin^{2N}(Jt/2)
+\cos(N\pi/2)\frac{|\sin^N(Jt)|}{2^{N-1}}\right]
\eeq
and the difference
\beq
\label{dfifference1}
\Delta
=\cos(N\pi/2)\frac{|\sin^N(Jt)|}{2^{N-1}}
=\cos(N\pi/2) p_e
.
\label{A_non}
\eeq
The periodic dynamics of the product $2|c_0||c_N|$, as defined in (\ref{non_int_exp}), 
is visualized for $N=2,10,20$ photons on the left-hand side of Fig. \ref{fig:1},
while the difference $\Delta$ is presented of the right-hand side of Fig. \ref{fig:1}. The evolution of the initial 
Fock state $|N,0\rangle$ is periodic with period $2\pi/J$, according to 
Eq. (\ref{non_int_exp}). For not too small $N$ 
the state $|N,0\rangle$ decays quickly, according to Eq. (\ref{decay0}), and after the time period $\pi/J$
a Fock state appears in the other cavity. This state disappears quickly again and the system returns after another period $\pi/J$ to the
initial Fock state, as visualized on the left-hand side of Fig. \ref{fig:2}. 
Thus, there is an anti-correlation effect: 
$c_0$ vanishes when $c_N$ becomes nonzero and vice versa. This effect increases with $N$, reflecting the fact that in the classical
limit $N\to\infty$ there is no entanglement. The probability for the creation of the PNS 
is related to $p_e(t)$, which decays exponentially with $N$, as given in Eq. (\ref{ent1}). 
Therefore, the maximal value $P_e$ is strongly suppressed. 
This is a consequence of the fact that for an increasing $N$ the photons disappear in 
the $(N+1)$--dimensional Hilbert space without contributing to the PNS.

\begin{figure}[t]
\begin{center}
\includegraphics[width=0.6\linewidth]{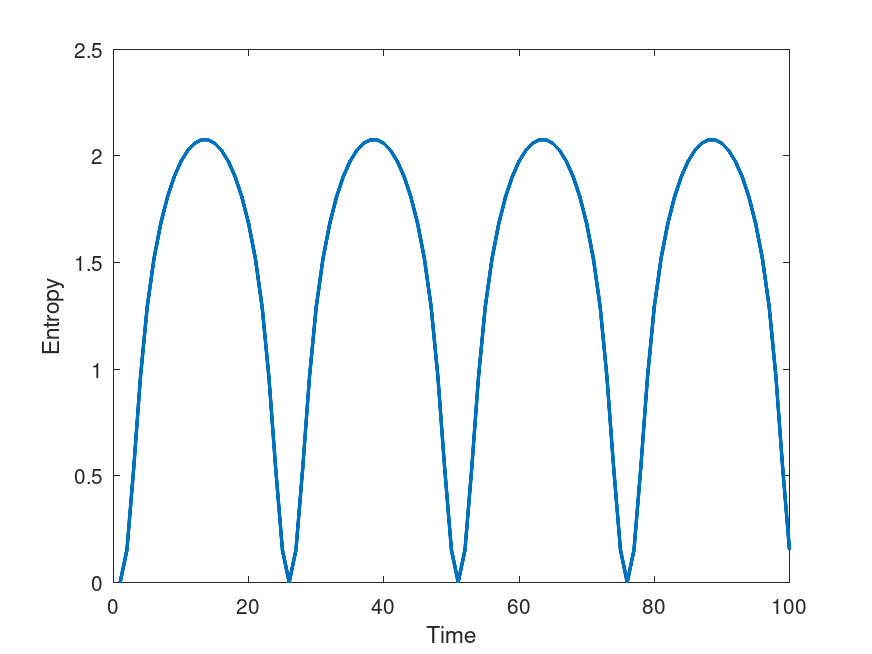}
\caption{
Time-periodic entanglement entropy for a unitary evolution in two coupled cavities with $N=20$ photons.
The time is given in units of $\hbar/J$.
}
\label{fig:4}
\end{center}
\end{figure} 

\subsection{Monitored evolution: measurement protocol}
\label{sect:monitoring}

We perform a projective measurement to find out if the quantum system is still in the 
pure initial state $|\Psi_0\rangle=|N,0\rangle$
or not after a unitary evolution over a time period $\tau$. If it is not, we allow another
unitary evolution over the time period $\tau$ and determine 
the probability that the quantum state is some given state $|\Psi\rangle$. This can be any
state of the Hilbert space, for instance, it can be the N00N state.
This procedure is repeated $m-1$ times ($m\ge 1$) and finally the probability 
$F_m$ for being the system in the state $|\Psi\rangle$ at time $m\tau$ is determined:
\beq
F_m=|\langle\Psi|T_\tau^{m-1}e^{-i \hat H_N \tau}|\Psi_0\rangle|^2
,
\eeq
where $T_\tau$ is the monitored evolution operator
\beq
\label{MEO}
T_\tau = e^{-i \hat H_N \tau} \left( {\bf 1} - |\Psi_0\rangle \langle \Psi_0 | \right)
.
\eeq
This provides a distribution of $\{F_m\}$ ($m\ge 1$) that characterizes the 
monitored evolution.
$T_\tau$ projects a state into the subspace orthogonal to $|\Psi_0\rangle$, while
the unitary operator describes a unitary evolution over a time period $\tau$.
This gives for the monitored evolution of the state $|\Psi_t\rangle$ over the time $t=m\tau$
the expression 
\beq
|\Psi_{m\tau}\rangle=T_\tau^{m-1}e^{-i \hat H_N \tau} |\Psi_0\rangle
.
\eeq

\subsection{Entanglement entropy}
\label{sect:ent_ent}

So far, we have studied the entanglement of two specific states, namely $|N,0\rangle$ and
$|0,N\rangle$. This is a local probe in a larger Hilbert space with dimension $N+1$.
As we have seen, the entanglement of these two states decreases strongly with $N$,
indicating that the entanglement of only two special states becomes less relevant as 
the size of the system increases. Alternatively, a probe of the entanglement of all 
between two complementary subspaces
is described by the EE, which in our specific case is the
entanglement between the two cavities.  It can be derived from the density operator,
as discussed subsequently. 

A quantum system is characterized by the density operator that reads in the 
presence of repeated projective measurements~\cite{liu23}
\beq
\label{density_operator00}
\rho^m(\tau)
=\frac{1}{{\cal N}}
T_\tau^{m-1}e^{-iH\tau}
|\psi_0\rangle\langle\psi_0|
e^{iH\tau}{T_\tau^\dagger}^{m-1}
\eeq
with the normalization 
${\cal N}=Tr[T_\tau^{m-1}e^{-iH\tau}|\psi_0\rangle\langle\psi_0|
e^{iH\tau}{T_\tau^\dagger}^{m-1}]$.
This density operator describes a quantum walk~\cite{PhysRevA.48.1687}, 
where after each time step $\tau$ a projective measurement $\Pi$ is applied. 
Although in general the projector $\Pi$ is independent of the initial state $|\psi_0\rangle$ and can be chosen
freely, in this paper we will focus on the case defined in Eq. (\ref{MEO}), where $\Pi$ 
projects onto the Hilbert 
space which is orthogonal to a given state $|\psi\rangle$ as 
$\Pi={\bf 1}-|\psi_0\rangle\langle\psi_0|$.
For the following discussion we start from the product space ${\cal H}_1\otimes{\cal H}_2$
of the two cavities with $n_1$ photons in the left cavity and $n_2$ photons in the right 
cavity. Then we assume that the Hamiltonian obeys particle conservation $n_1+n_2=N$,
which implies that it acts inside the Hilbert space that is spanned by the basis
$\{|N-n,n\rangle\}_{0\le n\le N}$. In this basis
the $(N+1)\times (N+1)$ density matrix elements read 
\beq
\rho^m_{N-n,n;N-n',n'}=\langle N-n,n|\rho^m(\tau)|N-n',n'\rangle
\eeq 
with $n,n'=0,\ldots ,N$. 
After summing over all basis states of ${\cal H}_2$ the reduced density matrix 
${\hat\rho}$ becomes an $(N+1)\times(N+1)$ diagonal matrix with elements
\beq
\label{reducedDM}
{\hat\rho}^m_{nn}
=\sum_{n'=0}^N\langle n,n'|\rho^m(\tau)|n,n'\rangle
=\langle n,N-n|\rho^m(\tau)|n,N-n\rangle
\]
\[
=\frac{1}{\cal N}\langle n,N-n|e^{-iH\tau}T_\tau^{m-1}
|\psi_0\rangle\langle\psi_0|T_\tau^{m-1}e^{iH\tau}|n,N-n\rangle
.
\eeq
With this expression for ${\hat\rho}^m_{nn}$ we can introduce the  
R\'enyi entropy~\cite{PhysRevX.8.041019} as a quantitative measure for the 
entanglement of the two Hilbert spaces ${\cal H}_1$ and ${\cal H}_2$:
\begin{equation}
\label{ent_entropy}
	{\cal S}_\alpha (\tau,N,m)
=\frac{1}{1-\alpha}\log_2 {\rm Tr}[{({\bar\rho}^m)^\alpha(\tau)}] 
.
\end{equation}
For the subsequent calculations we set $\alpha=2$, i.e., we will calculate 
${\cal S}_2(\tau,N,m)$.

\begin{figure}[t]
\begin{center}
\includegraphics[width=0.6\linewidth]{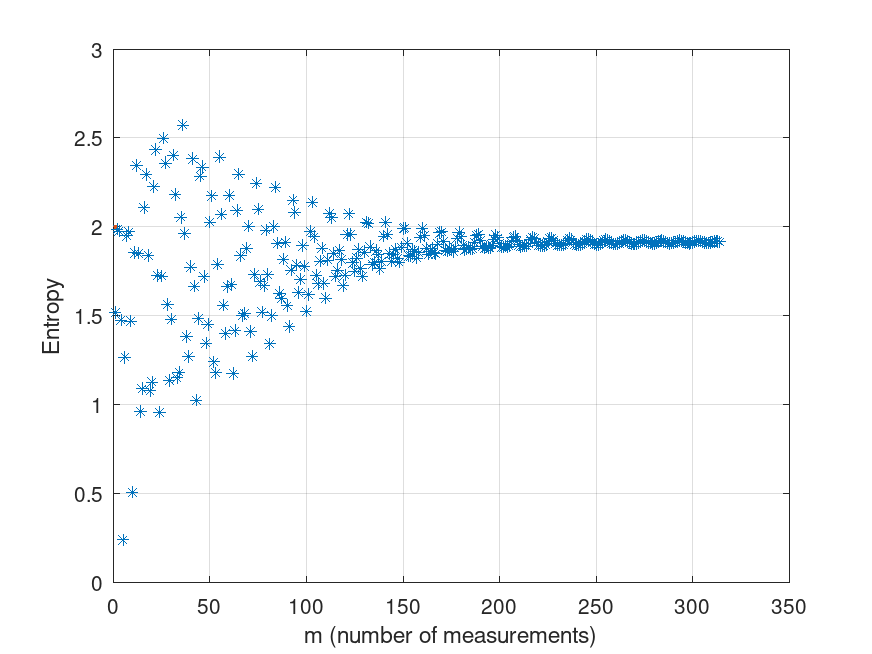}\\
\caption{
Entanglement entropy for $N=20$ photons in two coupled cavities with $J\tau/\hbar=\pi/10$.  
}
\label{fig:5}
\end{center}
\end{figure} 

\begin{figure}[t]
\begin{center}
\includegraphics[width=0.6\linewidth]{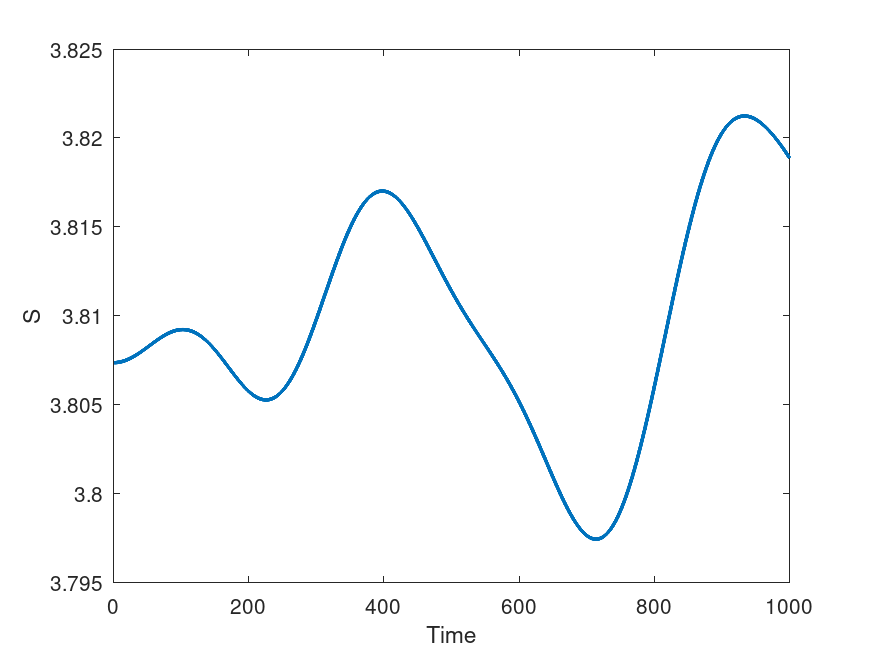}
\caption{
Entanglement entropy under a unitary evolution of photons, coupled to a single qubit
with coupling strength $\Omega/\omega=0.1$. The initial state is the superposition
$\rho_0=\sum_{n=1}^{15}|\d,n\rangle\langle\d,n|/15$.
The time is given in units of the inverse cavity frequency $1/\omega$.
}
\label{fig:6}
\end{center}
\end{figure} 

\section{Entanglement entropy of the Jaynes-Cummings model}
\label{sect:JC_model}

Besides the coupled cavities, another interesting example is the entanglement of 
cavity photons which couple to a single qubit, as it is presented by the 
Jaynes-Cummings model~\cite{jaynes63}. In this model, in contrast to the coupled
cavities, the number of photons is
not preserved because the qubit can absorb or emit a single photon.
It is characterized by a single-mode cavity frequency $\omega$ and
a two-level system (qubit) with transition frequency $\omega_a$.
The Hamiltonian reads in the rotating-wave approximation
(RWA)~\cite{HarocheRaimond}
\[
H = \hbar\left[ \omega a^\dagger a + \frac{\omega_a}{2} \sigma_z 
+\Omega (a^\dagger \sigma_- + a \sigma_+)\right]
,
\]
where $\hbar\Omega$ denotes the coupling strength between qubit and the cavity mode.
In this example the product Hilbert space is
${\cal H}_{\rm cavity}\otimes{\cal H}_{\rm qubit}$ with the basis 
$\{|\sigma,n\rangle\}_{\sigma=\u,\d; n=1,2,\ldots,N}$.
In this basis, the Hamiltonian matrix for this system reads
\beq
\hat H_N =
\pmatrix{
H_1 & 0 & \cdots & 0\cr
0 & H_2 & \ddots & \vdots\cr
\vdots & \ddots & \ddots & 0\cr
0 & \cdots & 0 & H_N\cr
}
\label{HN_full}
\eeq
with
\beq
H_{n}=
\frac{\hbar}{2}\pmatrix{
n\omega & \Omega\sqrt{n+1}\cr
 \Omega\sqrt{n+1} & (n+1)\omega \cr
}
.
\label{Hn_1q}
\eeq
Our objective is again to perform a monitored quantum walk.  That is, we start our 
system in some initial state 
\beq
\label{initial2}
|\psi_0\rangle = \sum_{n=1}^N c_n |\d,n\rangle\ \ {\rm with}\  \  \sum_{n=1}^N |c_n|^2=1
.
\label{psi_0_def}
\eeq
and allow it to evolve for a time interval $\tau$ unitarily. Then a projection 
$ {\bf 1} - |\psi_0\rangle \langle \psi_0 |$ is applied. This provides the time-dependent 
density matrix
\beq
{\hat\rho}_{m;\sigma n,\sigma\rq{}n\rq{}}=\frac{1}{\cal N}\sum_{n_0=1}^N|c_{n_0}|^2
\langle\sigma,n|e^{-i{\hat H}_N\tau}[T_\tau^\dagger]^{m-1}
|\d,n_0\rangle\langle\d,n_0|
T_\tau^{m-1}e^{i{\hat H}_N\tau}|\sigma\rq{},n'\rangle
,
\eeq
where $T_\tau$ is of the form given in Eq. (\ref{MEO}) and ${\cal N}$ is the normalization,
which is the trace of the matrix (cf. Eq. (\ref{density_operator00})). The density matrix
at time $t=0$ is $\rho_0=|\psi_0\rangle\langle\psi_0|$ with the initial state defined in
Eq. (\ref{initial2}).
The monitored quantum walk of the photons can be studied by the reduced density 
matrix that is obtained after tracing out the qubit:
\beq
{\bar\rho}^m_{nn'}=\sum_{\sigma=\u,\d}{\hat\rho}_{m;\sigma n,\sigma n\rq{}}
=\frac{1}{\cal N}\sum_{\sigma=\u,\d}\sum_{n_0=1}^N|c_{n_0}|^2
\langle\sigma,n|e^{-i{\hat H}_N\tau}[T_\tau^\dagger]^{m-1}
|\d,n_0\rangle\langle\d,n_0|
T_\tau^{m-1}e^{i{\hat H}_N\tau}|\sigma,n'\rangle
.
\eeq
This expression can be used to calculate the R\'enyi entropy in Eq. (\ref{ent_entropy})
as a quantitative measure for the entanglement between the qubit and the cavity photons.

\begin{figure}[t]
\begin{center}
\includegraphics[width=0.6\linewidth]{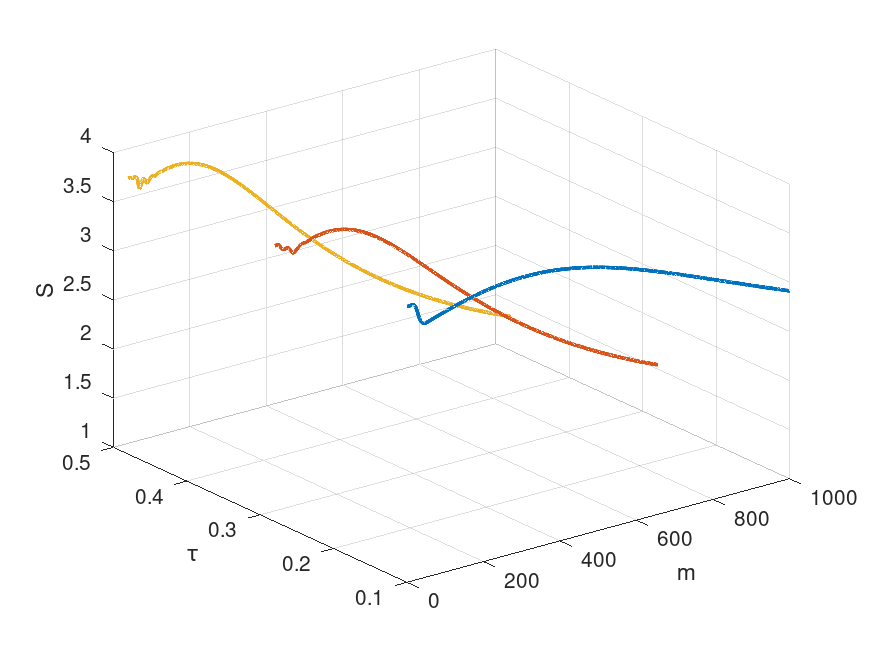}\\
\caption{
Entanglement entropy under a monitored evolution of photons, coupled to a single qubit
with coupling strength $\Omega/\omega=0.1$
for different time steps $\tau/\omega$ between measurements. The initial state is
the superposition $\rho_0=\sum_{n=1}^{15}|\d,n\rangle\langle\d,n|/15$.
}
\label{fig:7}
\end{center}
\end{figure} 

\section{Discussion of the results}
\label{sect:discussion}

The EE characterizes the entanglement of all states in the two cavities, while the fidelity 
and $\Delta$ describe the entanglement between two specific states only, namely 
$|N,0\rangle$ and $|0,N\rangle$. Thus, only the latter becomes less and less relevant 
when the Hilbert space is increased by an increasing number of photons. 

The  analytic results of the unitary evolution always lead to a periodic behavior whose
periodicity is determined by the coupling $J$ of the two cavities. For instance,
the return probability is $|c_0|^2=\cos^{2N} (Jt/2)$, while the transition probability 
$|c_N|^2= \sin^{2N} (Jt/2)$. This is also the case for the entanglement probability of a N00N
state, which reads $p_e=2|c_0c_N|=|\sin^N(Jt)|/2^{N-1}$, and for the difference 
$\Delta=\cos(N\pi/2)p_e$ of Eq. (\ref{difference1}). This is illustrated in Fig. \ref{fig:1}
for $N=2,10,20$ photons.
 
The return probability $|c_0|^2$ and the transition probability $|c_N|^2$ have a more complex
behavior under a monitored evolution. As illustrated for $N=100$ photons in Fig. \ref{fig:2},
the return to the initial state is significant only after 20 measurements, regardless of the time
$\tau$ between measurements. On the other hand, the transition to the state $|0,N\rangle$
is significant for less than 10 measurements, but depends on the time interval $\tau$.
The robustness of the return probabilities reflects the topological protection of the mean
return time under repeated projective measurements~\cite{gruenbaum13,Dhar2017}, which is also
robust even under random $\tau$ measurements~\cite{kessler21,ziegler21}. 

Entanglement in the form of a N00N state is illustrated in Fig. \ref{fig:3} for $N=10$ photons,
which indicates a decaying entanglement under monitored evolution: In the left column
the decay of the fidelity $|(\langle 0,N|+\langle N,0|)|\Psi_t\rangle|^2/2$ increases with
an increasing time step $\tau$. On the other hand, the difference $\Delta$ oscillates around
zero until it also decays. The oscillations disappear after 50 to 150 measurements, but details
depend on $\tau$ in a non-monotonic manner.

Switching to the probing of entanglement of many states, namely the entanglement between
states of the two cavities, we find a periodic behavior of the
EE under a unitary evolution, demonstrated in Fig.  \ref{fig:4} for $N=20$ photons
and the initial state $|N,0\rangle$. This reflects the periodicity of $c_0$ and $c_N$. The characteristic
periodicity does not depend on $N$, only on the coupling $J$ between the cavities. Moreover, the value of
the EE increases with the number of photons, because entanglement becomes more
complex with an increasing number of states.  
Under monitoring the EE has a stationary behavior of roughly 2 for two coupled cavities with 
$J\tau/\hbar=\pi/10$ (Fig. \ref{fig:5}). Since
the maximum of the entropy is obtained for an equipartion of $n$ elements,
it gives with $E_{\rm max}=\log_2 n\approx 2$ roughly $2$ independent partitions, reflecting the two
cavities (cf. Ref.~\cite{xavier18}).

The coupled cavities present a symmetric model with photon-number conservation. 
As an alternative we considered the coupling of the photons to a single qubit (e.g., a two-level atom)
which is inside the cavity. In this case the photon number fluctuates by one photon.
This system is described by the Jaynes-Cummings model and has been studied intensively
in theory and experiment~\cite{Cambetta,Carter,Viennot}. A typical coupling strength is 
$\Omega/\omega=0.1$.
For this case we find under a unitary evolution the behavior visualized in Fig. \ref{fig:6}.
The EE depends strongly on the initial density matrix, for which we choose
$\rho_0=\sum_{n=1}^{15}|\d,n\rangle\langle\d,n|/15$.
In contrast to the coupled cavities, after an initial regime with stronger fluctuations of the EE,
monitoring apparently makes the EE smoother (Fig. \ref{fig:7}), as a comparison with 
Fig. \ref{fig:6} illustrates. The decay on longer times indicates that the strongest entanglement is the
the result of the strong entanglement of the initial state. 
Since typically up to 40 measurements can be performed in a real quantum circuit until 
decoherence effects start to dominate~\cite{tornow23}, the decay of the EE may not be relevant in
practical applications.

\section{Conclusions}
\label{sect:conclusions}

In this Paper we analyzed the monitored dynamics of the Fock state of $N$
photons, initially located inside one cavity, coupled via an optical fiber to the second cavity,
as well as the coupling of a single qubit to $N$ photons. 
We monitored the entangled photonic $N00N$ state by projective measurements, 
repeated periodically with the time step $\tau$.

The transition and return amplitudes as well as the measure $\Delta$ for entanglement for two coupled 
cavities are periodic functions of time for a unitary evolution for different numbers of photons, where the
periodicity is determined by the coupling strength of the cavities. 
The probabilities of the return to the initial state$\left|N,0\right\rangle$ and the
transition to the state $\left|0,N\right\rangle$ for $N = 100$ as
functions of the number of measurements $m$ is not periodic but decays with increasing $m$.

The entangled state in the form of a N00N state is characterized by the fidelity and the difference $\Delta$ 
for two different N00N states are strong for a small number of measurements typically up to $m=100$,
but decays for larger $m$. The entanglement of all states between the two cavities, which is
characterized by the EE, has a periodic behavior. This reflects the fact that the system returns
periodically to the initial state $|N,0\rangle$ or transfers to the complementary state $|0,N\rangle$,
where the EE vanishes.
Under monitoring the EE fluctuates for a small numbers of measurements and can stabilize at a 
stationary value for larger $m$.

In addition to the system of two coupled cavities, we studied the monitored dynamics
of photons that are coupled to single qubit, based on the Jaynes-Cummings model.  
Then the EE is non-periodic already under unitary evolution. Monitoring results in a 
smoother behavior of the EE. 

Our results for the monitoring of the quantum evolution indicate a control over the
entanglement of $N$ photons. However, this is a complex approach that must be 
tuned for the specific system and for a specific outcome. In particular, the time $\tau$ 
between two successive measurements is an important parameter to establish a
useful protocol for a specific purpose.

\end{document}